\documentclass{article}
\usepackage{spconf_new,amsmath,graphicx}
\usepackage{bm}
\usepackage{algorithm}
\usepackage{algpseudocode}
\usepackage{comment}
\usepackage{multirow}
\usepackage{setspace}
\usepackage{url}
\usepackage{amssymb}
\usepackage {booktabs}
\usepackage{caption}
\usepackage{setspace}
\renewcommand{\thefootnote}{\fnsymbol{footnote}}
\renewcommand{\thefootnote}{\arabic{footnote}}
\title{NTT speaker diarization system for CHiME-7: multi-domain, multi-microphone End-to-end and vector clustering diarization}

\name{Naohiro Tawara, 
Marc Delcroix,
Atsushi Ando,
Atsunori Ogawa
}
\address{NTT Corporation, Japan}

\ninept

\begin{document}
\setlength{\baselineskip}{10.2pt} 
\maketitle
\begin{abstract}


This paper details our speaker diarization system designed for multi-domain, multi-microphone casual conversations. The proposed diarization pipeline uses weighted prediction error (WPE)-based dereverberation as a front end, then applies end-to-end neural diarization with vector clustering (EEND-VC) to each channel separately. It integrates the diarization result obtained from each channel using diarization output voting error reduction plus overlap (DOVER-LAP). To harness the knowledge from the target domain and results integrated across all channels, we apply self-supervised adaptation for each session by retraining the EEND-VC with pseudo-labels derived from DOVER-LAP. The proposed system was incorporated into NTT's submission for the distant automatic speech recognition task in the CHiME-7 challenge. Our system achieved 65 \% and 62 \% relative improvements on development and eval sets compared to the organizer-provided VC-based baseline diarization system, securing third place in diarization performance.

\end{abstract}

\begin{keywords}
speaker diarization, end-to-end neural diarization, vector clustering, CHiME-7 challenge
\end{keywords}

\section{Introduction}
\renewcommand{\thefootnote}{\fnsymbol{footnote}}
\footnotetext[0]{
© 2023 IEEE. Personal use of this material is permitted. Permission from IEEE must be obtained for all other uses, in any current or future media, including reprinting/republishing this material for advertising or promotional purposes, creating new collective works, for resale or redistribution to servers or lists, or reuse of any copyrighted component of this work in other works
}
\renewcommand{\thefootnote}{\arabic{footnote}}

In multi-party conversational speech analysis, speaker diarization plays an essential role in determining who speaks and when within the recordings.
Notably, it forms an important part of the preprocessing pipeline for the automatic speech recognition (ASR) system used in multi-party conversations~\cite{chime6,chime7}.

In natural multi-party settings, conversations are characterized by their large diversity, such as dynamic speaker turn-taking, highly overlapped speech, and a broad range of linguistic artifacts and speaking styles.
Conversations are usually recorded under varying conditions, such as background noises, reverberations, and multiple device configurations that include varying numbers of microphones, positions, and types of devices. 
The recently proposed CHiME-7 challenge~\cite{chime7} is a typical example of such challenging conditions consisting of natural conversations recorded with various distributed array microphones in diverse rooms.

The diversity in natural conversation makes the speaker diarization task extremely challenging, attracting considerable research effort~\cite{Park:22}.
Various diarization approaches have been proposed, including vector clustering (VC)~\cite{Sell:18}, target speaker voice activity detection (TS-VAD)~\cite{Medennikov:20}, end-to-end neural diarization (EEND)~\cite{Fujita:19, Horiguchi:20}, and EEND combined with VC (EEND-VC)~\cite{Kinoshita:21a, Kinoshita:21b}.

The VC-based approach is the most popular baseline system in many benchmarks. It accomplishes diarization by extracting speaker embeddings for short segments and then clustering them to assign speaker labels to each segment.
However, it assumes a single speaker in each segment, limiting the system performance on highly overlapping recordings. 

The recently introduced TS-VAD~\cite{Medennikov:20} addresses the issue of overlapping speech by utilizing a neural network that directly estimates speaker activities from a recording and speaker embeddings.
TS-VAD and its variants have achieved state-of-the-art performance in many diarization benchmarks~\cite{Medennikov:20, Cheng:23}.
However, TS-VAD requires an initial diarization to obtain speaker embeddings, and the VC-based approach is still required as pre-processing for TS-VAD.

EEND~\cite{Fujita:19, Horiguchi:20} is another widely-used neural diarization approach that directly estimates the speech activity for each speaker within a recording.
While EEND can estimate speech activity without requiring speaker embeddings, it faces challenges when applied to an arbitrarily large number of speakers or long-duration recordings~\cite{Horiguchi:20}.
EEND-VC~\cite{Kinoshita:21a, Kinoshita:21b} is a hybrid approach of VC and EEND that was introduced to combine the strengths of both frameworks. 
It first performs EEND on the speech segment to estimate the speaker activities and embeddings. Then, it performs VC on the estimated speaker embeddings to stitch together the speaker activities of the same speaker across different segments. 
EEND-VC can handle overlapping speech like EEND while accommodating an arbitrary number of speakers and long recordings like VC.
The primary limitation of neural network-based approaches, including EEND-VC, is their tendency to overfit the training data, making it challenging to generalize to new conditions that may involve varying noise and recording devices.
Thus, fine-tuning the model using data from the target domain is essential~\cite{Fujita:19}. 
However, obtaining sufficient data in the target domain is not always feasible.

To tackle these issues, we introduce a novel EEND-VC-based speaker diarization system designed for highly overlapping, multi-domain, and multi-microphone conditions.
Our system starts by applying multi-channel weighted prediction error (WPE)-based dereverberation~\cite{Nakatani:10} and then applies EEND-VC to individual channels. 
Then, the system integrates the dirization result with diarization output voting error reduction plus overlap (DOVER-LAP)~\cite{Raj:21}. 
Finally, the system retrains the EEND-VC model using the channel-integrated diarization result in a self-supervised adaptation (SSA)~\cite{Takashima:21} manner and conducts another round of estimation with the adapted models.
Our proposed system is incorporated into NTT’s submission for the distant automatic speech recognition (DASR) task in the CHiME-7 challenge, outperforming the challenge baseline with a relative diarization error rate (DER) improvement of 65 \%. 
It achieved the third position in the CHiME-7 challenge, showing the potential for EEND-VC.

This paper describes our system in detail and provides a thorough analysis. 
The highlights of our findings are three-fold.
%
%
First, our analysis reveals that the embeddings from EEND-VC suffer from a large speaker confusion on highly overlapping segments. However, we also show that this issue can be mitigated by incorporating the pretrained speaker embedding model, such as ECAPA-TDNN~\cite{Desplanques:20}.
Second, we show that the performance of EEND-VC is highly dependent on the segment length. 
While sufficiently long segments (e.g., 80 seconds) are necessary to capture speaker characteristics, overly long segments can lead to intra-segment speaker permutation errors in activity estimation.
We experimentally show that constrained clustering plays an important role in reducing the influence of this permutation error.
Third, we show the effectiveness of DOVER-LAP-based channel integration along with the SSA approach, indicating that this not only refines the results on individual channels but also elevates the overall system performance. 

\section{System description}
\label{sec.exp}

\begin{figure}[tb]
 \centering
 \includegraphics[width=0.3\textwidth]{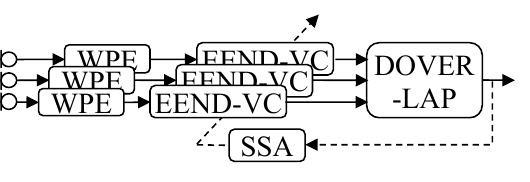}
 \vspace{-5mm}
 \caption{Proposed multi-channel speaker diarization system.}
 \label{fig:system_overview}
\vspace{-2mm}
\end{figure}

\begin{figure}[tb]
 \centering
 \includegraphics[width=0.47\textwidth]{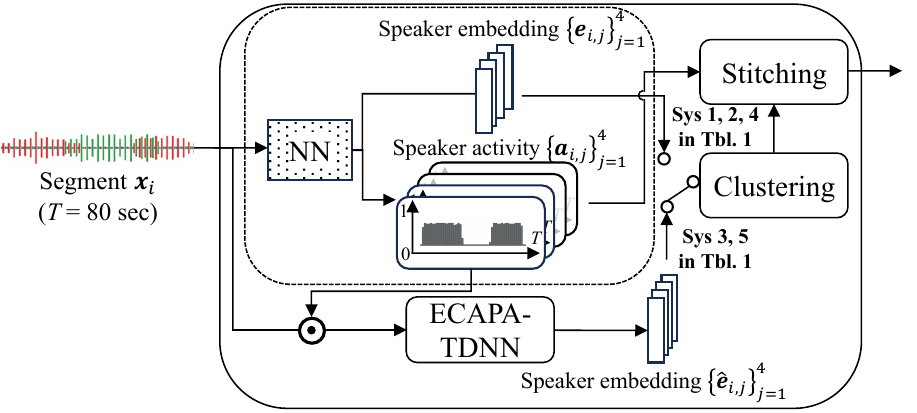}
 \vspace{-2mm}
 \caption{Schematic diagram of EEND-VC with ECAPA-TDNN.
 The part enclosed by dotted lines corresponds to the original EEND-VC.
 Only NN parameters are retrained in SSA.}
 \label{fig:eend-vc}
\vspace{-2mm}
\end{figure}

Figure~\ref{fig:system_overview} presents the overview of our diarization system. 
First, it applies multi-channel WPE-based dereverberation~\cite{Nakatani:10} and EEND-VC~\cite{Fujita:19} to each channel separately. 
Then, it combines the results using DOVER-LAP~\cite{Raj:21}. 
Finally, it performs SSA~\cite{Takashima:21} on each session and each microphone to retrain the EEND-VC model using labels obtained by DOVER-LAP.
We provide details of each module in the following subsections.

\subsection{EEND-VC-based diarization}
\label{subsec:eendvc}

Figure~\ref{fig:eend-vc} shows the EEND-VC architecture.
It first divides the entire recordings into segments $\boldsymbol{x}_{i}\in{\mathcal R}^{T}$, and then estimates both local speaker activities $\{\boldsymbol{a}_{i,j}\in[0,1]^{T}\}_{j=1}^{S}$ and the local speaker embeddings
$\boldsymbol{e}_{i} = \{\boldsymbol{e}_{i,j} \in {\mathcal R}^{D}\}_{j=1}^{S}$ for each segment as 
\begin{equation}
\{\boldsymbol{a}_{i}, \boldsymbol{e}_{i}\} = {\rm NN}(\boldsymbol{x}_{i}),
\end{equation}
where ${\rm NN}$ denotes the encoder in EEND-VC, and $S$, $T$, and $D$ denote the number of local speakers, segment size, and embedding dimensions, respectively. 

In the upcoming experimental section, we will show that the segment size $T$ significantly impacts the system's overall performance. Notably, utilizing long sentences improves performance by providing a greater non-overlapping region within each segment, yielding more informative speaker embeddings.
Our framework assumes that the maximum number of speakers throughout the recording is predetermined and does not exceed the number of local speakers $S$, as stipulated by the CHiME-7 challenge regulation.
This enables us to process longer segments than that in conventional studies~\cite{Kinoshita:21a, Kinoshita:21b}. 
We use a segment size of 80 seconds.

In addition, our experiments will also observe that the EEND-VC system suffers from a large speaker confusion on highly overlapped segments.
To tackle this issue, we incorporate a pre-trained ECAPA-TDNN model~\cite{Desplanques:20} into the EEND-VC system.
Specifically, we employed ECAPA-TDNN to extract the speaker embeddings $\hat{\boldsymbol{e}}_{i, j}$ for $j$-th  speaker using the speech activity estimated by EEND-VC as
\begin{equation}
    \hat{\boldsymbol{e}}_{i, j}= {\rm ECAPA}(\boldsymbol{a}_{i,j} \odot \boldsymbol{x}_{i}).
\end{equation}

Finally, EEND-VC performs clustering of speaker embeddings to stitch the segments together to form the diarization results.
In the EEND-VC-based approach, each segment may contain multiple speaker embeddings, which should be of different speakers.
Accordingly, we need to introduce constraints during the clustering process: embeddings from the same segment must not be grouped into the same cluster.
Cop-Kmeans clustering~\cite{Wagstaff:01} is suitable for this purpose, as it can strictly impose this constraint.
However, estimating the number of speakers with this method is challenging.
Instead, we adopt the following two-step clustering procedure.
First, we apply constraint agglomerative hierarchical clustering (cAHC)~\cite{Davidson:09} to estimate the number of speakers under a soft constraint. 
We estimate the number of speakers as the minimum between the cluster count determined by cAHC and a predefined maximum number of speakers. We then perform a second round of clustering using COP-K-means, which allows for stricter constraints, with the number of clusters fixed to the estimated number of speakers.

\subsection{Multi-channel integration}
\label{subsec:doverlap}

When multiple channels are available, several channel integration approaches can be employed. 
For example, exploiting fine spatial information such as spatial feature clustering or augmenting EEND with multi-channel features \cite{Ishiguro:11, Horiguchi:22} can provide important clues to discriminate speakers. However, it requires speaker tracking when the speakers frequently move and may also be sensitive to the array topology. 
An alternative is the late fusion approach, which integrates the diarization results estimated on each channel independently~\cite {Boeddeker:23}. Late fusion cannot exploit fine spatial information but may be more robust to speaker movements and changes in the array topology because it processes each channel independently.

We opt for the latter option in this paper and perform the diarization on each channel independently using EEND-VC described in Section \ref{subsec:eendvc}. Then, we combine the diarization results of all the available channels using DOVER-LAP~\cite{Raj:21}.
For DOVER-LAP, we use the Hungarian algorithm to find the best permutation between the diarization results of the different channels and then combine them using a voting scheme.
Some channels may encounter difficulties such as obstacles and failure of recording. This voting scheme can potentially mitigate the impact of these outlier results.
\subsection{Semi-supervised adaptation }

To obtain further improvement, we adapt the EEND-VC to the individual channel of the target session with SSA as proposed in \cite{Takashima:21}.
Specifically, we first apply our systems to test data and obtain the diarization result, which will serve as pseudo-labels for adaptation. Here, we use DOVER-LAP to combine all microphones' results as described in section \ref{subsec:doverlap}, which provides more reliable pseudo-labels.
Then, we retrain the model for each channel in each session using each test data using the estimated label.
We repeat this process multiple times to obtain further improvement.

\begin{table}[t]
  \caption{Diarization results on the {\bf DEV} set in terms of confusion (CF), false alarm (FA), missed (MI), and DER computed with dscore toolkit~\cite{dscore} with a collar of 0.25 sec. The best DERs for each scenario are highlighted.}
  \vspace{-3mm}
  \label{tab:diarization_results}
  \centering
  \resizebox{\linewidth}{!}{
  \begin{tabular}{@{}l@{\hspace{0.2cm}}lc@{\hspace{0.2cm}}c@{\hspace{0.2cm}}c@{\hspace{0.2cm}}cc@{\hspace{0.2cm}}c@{\hspace{0.2cm}}c@{\hspace{0.2cm}}cc@{\hspace{0.2cm}}c@{\hspace{0.2cm}}c@{\hspace{0.2cm}}cc@{\hspace{0.2cm}}cc@{}}
    \toprule
    & & \multicolumn{4}{c}{CHiME-6} & \multicolumn{4}{c}{DiPCo (S26, S29)}& \multicolumn{4}{c}{Mixer 6} & Macro\\ 
    ID &Model & CF & FA & MI &DER & CF & FA & MI &DER & CF & FA & MI &DER &  DER \\
    \midrule
  & Baseline        & 14.5 & 3.2 & 22.3 & 40.0 & 13.0 & 4.7 & 12.0 & 29.8 & 1.7 & 1.0 & 13.8 & 16.6 & 28.8 \\
\midrule
Sys1 & EEND-VC              & 15.1 & 3.6 & 17.9 & 36.6 & 6.9 & 4.1 & 9.7 & 20.7 & 0.6 & 1.7 & 8.0 & 10.3 & 22.5 \\
Sys2 & Sys1 + WPE           & 17.9 & 3.6 & 16.6 & 38.0 & 7.0 & 3.3 & 9.1 & 19.5 & 0.2 & 1.8 & 7.8 & 9.8  & 22.4  \\
Sys3 & Sys2 w/ ECAPA   & 8.9  & 3.8 & 13.3 & 28.9 & 5.6 & 3.8 & 9.2 & 18.5 & 0.2 & 1.5 & 4.2 & 9.9  & 19.1 \\
\midrule
Sys4 & Sys3 + SSA           & 10.7 & 4.5 & 15.5 & 30.7 & 5.8 & 3.1 & 9.4 & 18.2 & 0.2 & 1.9 & 7.7 & \underline{\bf 9.7}  & 19.5 \\
Sys5 & Sys4 w/ ECAPA   & 7.7  & 5.1 & 14.9 & \underline{\bf 27.7} & 5.2 & 3.1 & 9.6 & \underline{\bf 17.9} & 0.8 & 2.0 & 7.6 & 10.4 & \underline{\bf 18.7} \\
     \bottomrule
  \end{tabular}
}
  \vspace{-0mm}
\end{table}
\section{Experiments}
\label{sec.exp}

We evaluated our system using the CHiME-7 dataset~\cite{chime7}, following the DASR task regulation.

\subsection{Data resource}
The CHiME-7 dataset comprises three subsets taken from CHiME-6~\cite{chime6}, DiPCo~\cite{dipco}, and Mixer 6 Speech~\cite{mixer6}.
Each subset encompasses a specific scenario, reflecting potential real-world applications. 
These scenarios differ in array topologies: 6 far-field Kinect 4-mic microphones arrays (CHiME-6), five far-field devices with a 7-mic circular array (DiPCo), or a heterogeneous combination of 14 microphones (Mixer 6), as well as in diverse situations, such as dinner parties (CHiME-6 and DiPCo) and interviews (Mixer 6). 
The number of speakers in each session is not provided, but we can use the information that the maximum number of speakers in each session is up to four.
The use of external data is restricted to a limited extent, with allowances made for specific external data sources and pre-trained models.
More detailed data description and challenge regulations can be found in~\cite{chime7}.

\subsection{Experimental Setup}

\begin{table}[t]
  \caption{Diarization performance comparison with top-5 participants on {\bf EVAL} set\protect\footnotemark[1]. The results are sorted with Macro DER, highlighting each scenario's best DERs. 
  }
  \vspace{-3mm}
  \label{tab:all_results}
  \centering
  \scriptsize
  \setlength{\tabcolsep}{.6mm}

  \begin{tabular}{lccccc}
    \toprule
    & Main & CHiME-6 & DiPCO& Mixer 6 & Macro\\ 
    Team & Approach & DER & DER & DER & DER\\
    \midrule
USTC-NERCSLIP   & TS-VAD  &          \underline{\bf 25.1}  & \underline{\bf 16.4} & 6.1   & \underline{\bf 15.9} \\
IOA-CAS-Speech  & TS-VAD             & 27.3  & 22.4 & 7.3                 & 19.0\\
NTT (Sys5)      & Clustering    & 31.3  & 21.1 & \underline{\bf 5.9} & 19.4 \\
NTT (Sys3)      & Clustering    & 32.3  & 22.6 & 5.8 & 20.2 \\
NTT (Sys2)      & Clustering    & 39.1  & 22.7 & 5.8 & 22.5 \\
Univ. Cambridge & Clustering         & 48.2  &  25.6  &  10.3  &  28.0  \\
Baseline        & Clustering        & 56.3  &  27.9  &  9.3  &  31.2  \\
     \bottomrule
  \end{tabular}
  \vspace{-3mm}
\end{table}
We used the pre-trained WavLM-large to obtain the input speech features as proposed in~\cite{WavLM}.
The outputs from all WavLM's transformer layers were averaged using learnable weights, yielding 1024-dimensional input speech features.
We used the EEND-VC model similar to that in~\cite{WavLM}.
It consisted of six-stacked Transformer-encoder blocks with eight 256-dimensional attention heads.
We projected the encoder’s output with a linear layer into four output streams, each consisting of the frame-by-frame speaker activity binary decisions and the 256-dimensional speaker embedding.


We adopted a two-stage approach to train the EEND-VC model.
First, we trained the model using simulated multi-talker recordings of up to four speakers. 
These simulated mixtures were generated following the method described in~\cite{Fujita:19}, using data from LibriSpeech~\cite{Librispeech}, noises from MUSAN~\cite{MUSAN}, and room impulse responses (RIRs) from SLR28~\cite{rir_data}. 
Each mixture contained up to four speakers and had an average silence duration of two seconds ($\beta=2$) between utterances from the same speaker. 
We obtained 100,000 speech mixtures from this process, totaling 12,685 hours and 2,338 speakers.
We then fine-tuned the model using recordings from randomly selected channels in the CHiME-6 and Mixer~6 training sets and a part of the DiPCo development set.
We used the Mixer~6 interview training set, labeled only for interviewees. For interviewer labels, we applied our early diarization system fine-tuned on CHiME-6 to lapel recordings. Our training also included sessions S28, S33, and S34 from the DiPCo development set.
The total fine-tuning data was 82 recordings, 80.3 hours, and 114 speakers.

For the first training stage, we fixed the WavLM parameters and trained the model for 25 epochs with a learning rate of $10^{-3}$, a batch size of 2048, and a segment size of 15 sec. 
We used the Adam optimizer with 25000 warm-up steps.
In the second stage, we fine-tuned the whole model on segments of 80 sec for three epochs with a learning rate of $10^{-5}$ and a batch size of one. 
For SSA, we retrained the model using the labels obtained by the DOVER-LAP for each session and microphone independently for two epochs with a learning rate of $10^{-5}$. We repeated the SSA process twice.

During inference, we applied WPE~\cite{Nakatani:10} to remove the reverberation before applying EEND-VC.
We computed the WPE filter on the short-time Fourier transform spectrum with a 64 ms window and a 16 ms shift. 
We set the prediction delay and the filter length to two and one, respectively.
For speaker embeddings extraction, we used SpeechBrain's ECAPA-TDNN~\cite{Speechbrain, Desplanques:20} pretrained on 7,000+ speakers' 2,000+ hours utterances included in VoxCeleb1\&2~\cite{Nagrani:19}.

\subsection{Results}

Table~\ref{tab:diarization_results} shows the dirization performance on the DEV set in terms of speaker confusion (CF), false alarm (FA), missed (MI), and DER.
The results for each scenario are reported separately, and the macro-averaged DER across all scenarios is presented.

The first row in the table represents the result from the CHiME-7 baseline; the others are for our proposed pipeline.
Systems 1, 2, and 4 performed the clustering on embeddings obtained by EEND-VC depicted as $\boldsymbol{e}_{i}$ in Fig.~\ref{fig:eend-vc}. 
The other systems relied on ECAPA-TDNN embeddings
depicted as $\hat{\boldsymbol{e}}_{i}$ in Fig.~\ref{fig:eend-vc}.
Notably, except System 1, all our systems used the WPE-based dereverberation front-end.

The comparison between Systems 1 and 2 indicates that WPE contributed to reducing MI and FA while slightly deteriorating CF.
This was probably because WPE sharpened segment boundaries but introduced minor distortions. 
However, this slight deterioration in CF is inconsequential because subsequent ECAPA-TDNN-based systems solely leverage segment information derived from EEND-VC.
The comparison between Systems 2 and 3 demonstrates that the Macro DER was reduced by 3.3 points by combining ECAPA-TDNN embeddings with the EEND-VC framework. 
Particularly, a significant reduction of CF was observed in CHiME-6. 
This result indicates the limitation of the EEND-VC embeddings, probably due to the limited speaker variability in the training set.
System~4 shows the result of the adapted model using SSA by retraining the EEND-VC (System 2) using labels derived from System 3.
Compared with the pre-SSA (System~2), SSA substantially decreased CF, improving DER across all scenarios. 
Moreover, combining ECAPA-TDNN with System~4 provides a further improvement, achieving the lowest DER in all scenarios except for the Mixer~6.

Table~\ref{tab:all_results} lists the diarization results on the evaluation set, including the top-5 teams\footnotemark[1]. 
Our system ranked third in CHiME-6, second in DiPCO, and achieved the first-place for Mixer6. 
Regarding the macro DER, our system achieved third place.
Notably, our system delivered the finest performance among all VC-based systems.

\subsection{Analysis}
\begin{figure}[t]
\begin{center}
\begin{tabular}{ccc}
\hspace{-5pt}\includegraphics[width=.28\linewidth]{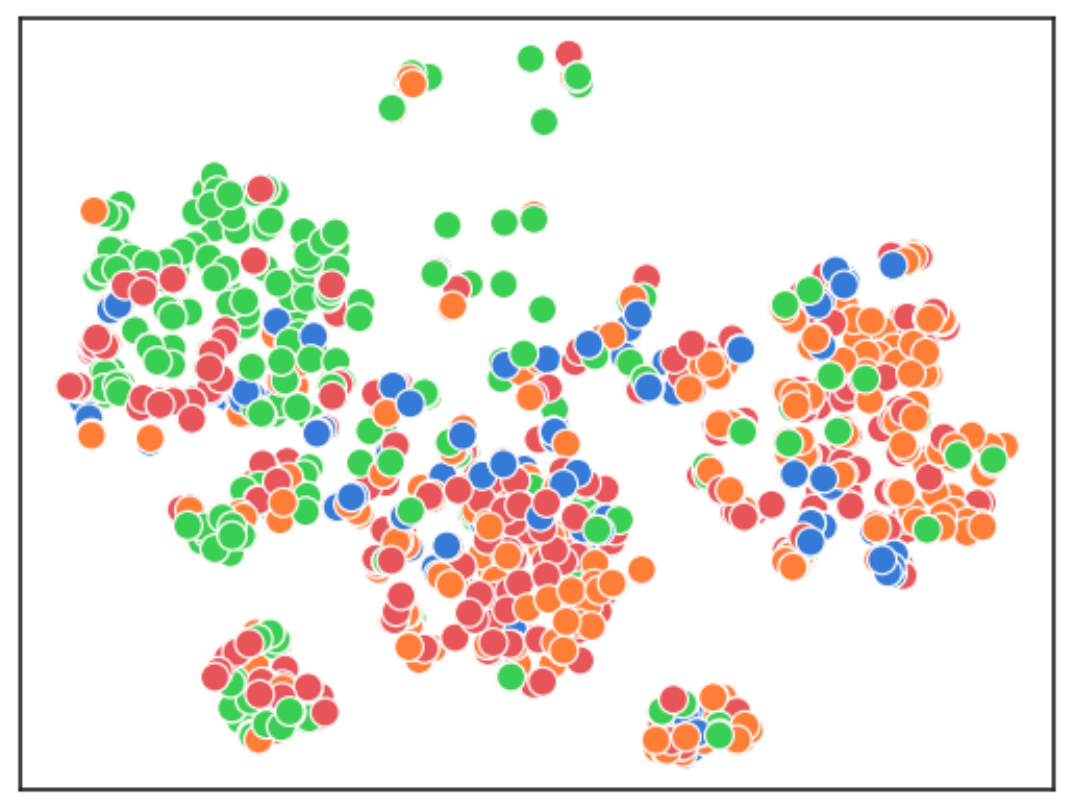} &
\hspace{-10pt}\includegraphics[width=.28\linewidth]{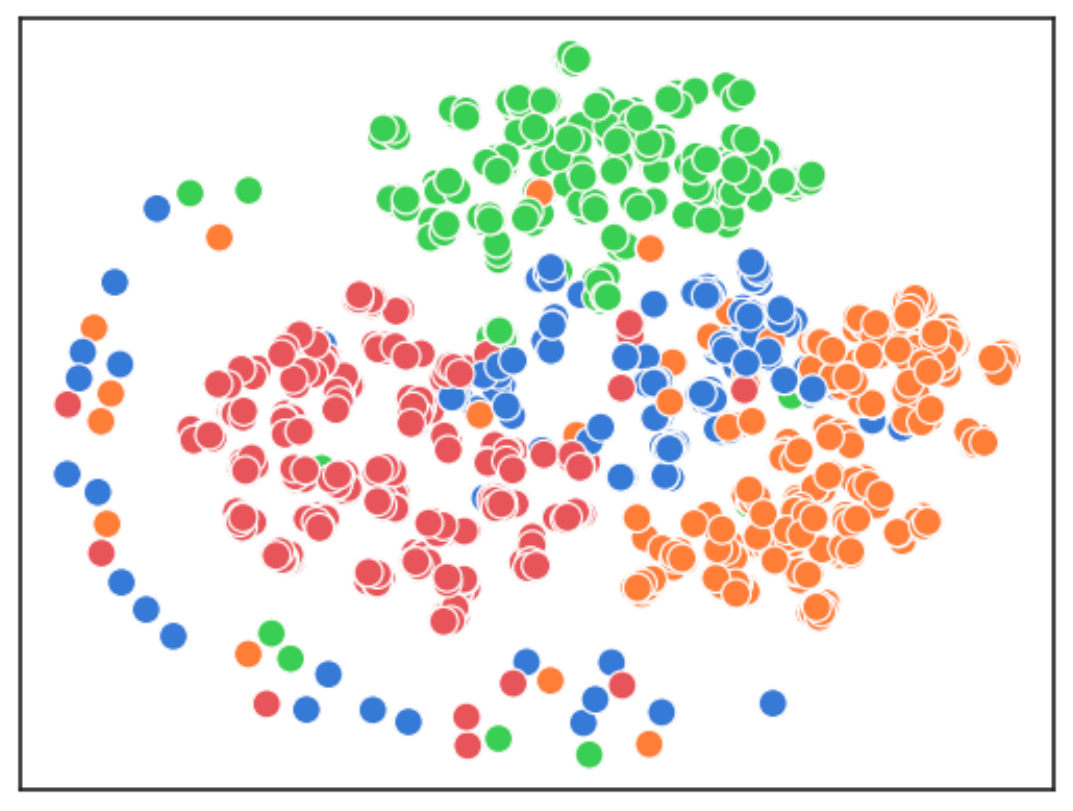} &
\hspace{-10pt}\includegraphics[width=.28\linewidth]{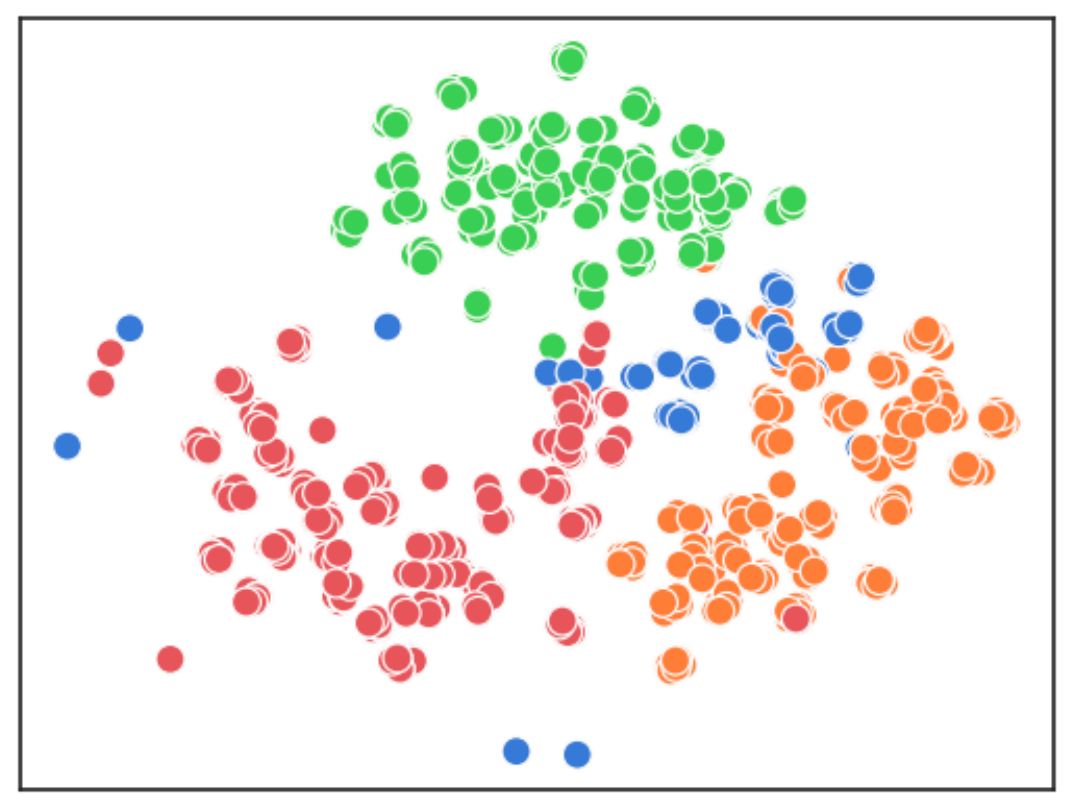}\\
\hspace{-5pt} (a) EEND-VC & 
\hspace{-10pt} (b) ECAPA &
\hspace{-10pt}(c) ECAPA + SSA \\
\end{tabular}
\end{center}
\vspace{-5mm}
\caption{
Visualization of embeddings obtained by our systems on S29.
Each plot corresponds to the embeddings colored by speaker labels.
}
\label{fig:embeddings}
\vspace{-1mm}
\end{figure}
\footnotetext[1]{
Our result shows a slight improvement over the submitted model that was optimized for ASR tasks.
All results can be found in \url{https://www.chimechallenge.org/current/task1/results}.
}
Figure~\ref{fig:embeddings} provides a t-SNE visualization~\cite{tsne} of the embeddings generated by our systems. 
Individual plots in the figure correspond to the embeddings of segments extracted by EEND-VC or ECAPA-TDNN. 
The embeddings are colored based on the speaker labels determined with oracle clustering.
This figure indicates that ECAPA-TDNN embeddings exhibit a more distinct separation among speakers than those derived from EEND-VC. 
This implies that the ECAPA-TDNN technique provides a better feature representation for speaker diarization.
This figure also indicates that superior embeddings were obtained by applying SSA.

\begin{figure}[tb]
 \centering
 \hspace{-5mm}\includegraphics[width=0.38\textwidth]{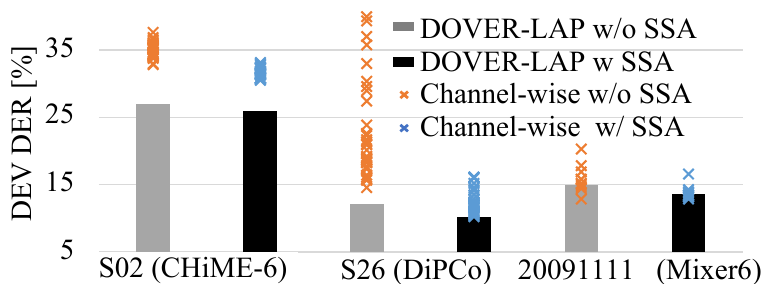}
 \vspace{-1mm}
 \caption{
 Session- and channel-wise diarization results for the ECAPA-TDNN systems, both with and without SSA. 
 Individual plots represent channel-wise results, while the bars the DOVER-LAP results.
 }
 \label{fig:session_result}
 \vspace{-3mm}
\end{figure}
To analyze the effectiveness of DOVER-LAP and SSA, we depict the channel-wise diarization results of randomly selected sessions derived from ECAPA-TDNN systems with and without SSA in Fig.~\ref{fig:session_result}. 
Individual plots represent channel-wise results, while the bars show the DOVER-LAP results.
These results indicate that the DERs were notably worse on specific channels for the model before applying SSA.
Specifically, the channel-wise results for specific channels in session S26 were noticeably poorer. This is likely because these channels are located near obstacles, such as a TV, making diarization extremely challenging.
Despite this, these channel-specific errors were greatly reduced upon applying SSA, and DOVER-LAP improved these results further.
This improvement is probably attributed to the labels used in SSA, which were derived from the channel-integrated results. Consequently, adapting the model to each channel reduced channel-specific errors by leveraging the insights from other channels' results.

\begin{figure}[tb]
 \centering
 \includegraphics[width=0.36\textwidth]{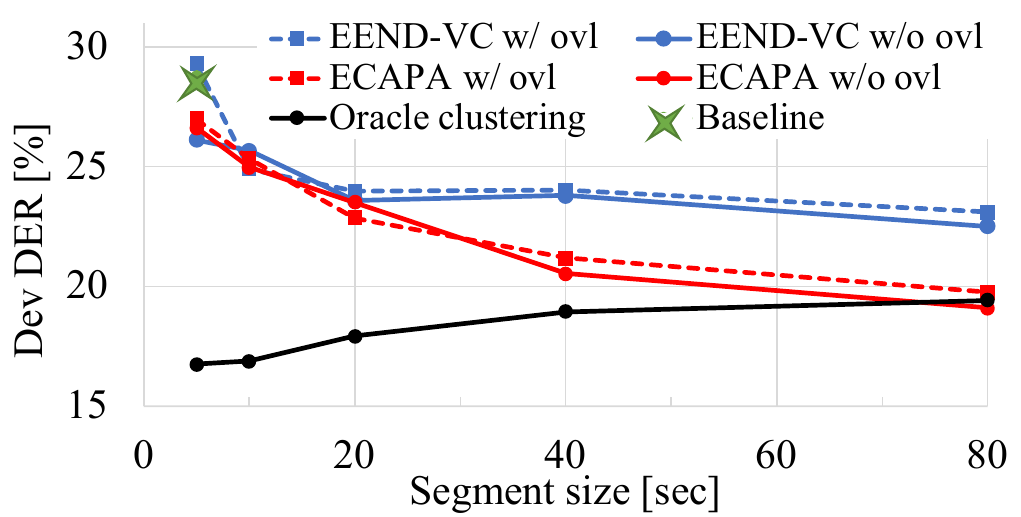}
 \vspace{-2mm}
 \caption{The macro DER on the DEV set obtained based on EEND-VC and ECAPA-TDNN embeddings, plotted against segment size.}
 \label{fig:segment_size_result}
 \vspace{-1mm}
\end{figure}

\begin{table}[t]
\centering
\renewcommand{\baselinestretch}{}
\caption{
Performance comparison of different clustering algorithms in the ECAPA-TDNN system (System 3). The column \#unconstraint denotes the segment count not satisfying the clustering constraint.
}
\label{tab:clusteinrg_result}
\scriptsize
\vspace{-7mm}
\begin{center}
\begin{tabular}{lcc}
\toprule
Algorithm       &   Macro DER & \#unconstraint \\ 
\midrule
AHC & 19.8 & 3405  \\
cAHC~\cite{Davidson:09} & 19.7 & 3060  \\
\midrule
Kmeans & 20.4 & 3849  \\
COP-Kmeans~\cite{Wagstaff:01} & {\bf 19.1} & {\bf 0} \\
\bottomrule
\vspace{-8mm}
\end{tabular}
\end{center}
\end{table}
Figure~\ref{fig:segment_size_result} illustrates the relationship between segment size and the macro DER of both EEND-VC- and ECAPA-TDNN-based systems before applying SSA (Systems 2 and 3).
The red, blue, and black lines represent the results of clustering based on EEND-VC, ECAPA-TDNN, and oracle clustering, respectively. 
The solid lines depict the results obtained by extracting embeddings excluding overlapped frames, while the dashed lines depict the results that included these frames.
This result indicates that longer segments yield superior performance. 
At the same time, longer segments deteriorated the performance of the oracle clustering.
The primary reason for this discrepancy is that although longer segments provide richer speaker information, they also tend to cause inner-segment speaker permutation errors.  
Eventually, the optimal performance was achieved when extracting embeddings from 80-second segments without overlaps.
Notably, clustering with ECAPA-TDNN achieved a comparative performance to the oracle clustering.

The gap in performance between results derived from clustering and those from oracle clustering on short segments, shown in Fig.~\ref{fig:segment_size_result}, suggests an intrinsic limitation of the VC-based approach.
To examine this issue, we analyzed the local speaker activities identified by our system. We observed inner-segment permutation errors, where the activities of different speakers were mistakenly recognized as those of a single speaker.
This inner-segment speaker permutation error harms the clustering process because the same speaker's activity appeared in different streams of the local speaker activities.
The constraint in clustering introduced in Section 2 is expected to ease this problem.
To validate this, we present the macro DER and the count of samples that did not meet the constraint using different clustering algorithms, including AHC, constraint AHC, Kmeans, and COP-Kmeans, in Table~\ref{tab:clusteinrg_result}. 
This result indicates that only COP-Kmeans could force all segments to satisfy the constraint, delivering the best performance. 
This underscores the importance of hard clustering constraints in mitigating the detrimental impact of permutation errors.

\section{Conclusion}
This paper described our EEND-VC-based speaker diarization system designed for highly overlapping, multi-domain, and multi-microphone conditions.
Our system achieved third place in the CHiME-7 challenge, providing some findings important for the VC-based approach.
In future works, we plan to investigate applying the more sophisticated clustering algorithm such as multi-stream VBx~\cite{Delcroix:23}. 
We also plan to incorporate more sophisticated structures such as memory-aware embedding~\cite{He:23} into EEND-VC to obtain better activity and embeddings.

\bibliographystyle{IEEEbib-abbrev}
\begin{spacing}{.88} 
\bibliography{ref}
\end{spacing}

\end{document}